\documentclass[preprint,12pt]{aastex}
\usepackage{psfig}
\newcommand{\beq}{\begin{equation}}
\newcommand{\eeq}{\end{equation}}
\newcommand{\bea}{\begin{eqnarray}}
\newcommand{\eea}{\end{eqnarray}}
\newcommand{\bean}{\begin{eqnarray*}}
\newcommand{\eean}{\end{eqnarray*}}
\newcommand{\ba}{\begin{array}}
\newcommand{\ea}{\end{array}}
\newcommand{\bml}{\begin{mathletters}}
\newcommand{\eml}{\end{mathletters}}
\newcommand{\rem}[1]{{ }}
\bibpunct[,]{(}{)}{;}{a}{}{,}
\begin{document}
\title{On the Lack of Type I X-ray Bursts in Black Hole X-ray Binaries:
Evidence for the Event Horizon?}

\author{Ramesh Narayan\altaffilmark{1,2} \& Jeremy S. Heyl\altaffilmark{2,3}}

\altaffiltext{1}{Institute for Advanced Study, Princeton, NJ 08540}

\altaffiltext{2}{Harvard-Smithsonian Center for Astrophysics, Cambridge, 
MA 02138; rnarayan@cfa.harvard.edu; jheyl@cfa.harvard.edu}

\altaffiltext{3}{Chandra Fellow} 

\begin{abstract}

Type I X-ray bursts are very common in neutron star X-ray binaries,
but no Type I burst has been seen in the dozen or so binaries in which
the accreting compact star is too massive to be a neutron star and
therefore is identified as a black hole candidate.  We have carried
out a global linear stability analysis of the accumulating fuel on the
surface of a compact star to identify the conditions under which
thermonuclear bursts are triggered.  Our analysis, which improves on
previous calculations, reproduces the gross observational trends of
bursts in neutron star systems.  It further shows that, if black hole
candidates have surfaces, they would very likely exhibit instabilities
similar to those that lead to Type~I bursts on neutron stars.  The
lack of bursts in black hole candidates is thus significant, and
indicates that these objects have event horizons.  We discuss possible
caveats to this conclusion.
\end{abstract} \keywords{accretion --- black hole physics --- X-rays:
binaries, bursts}

\section{Introduction}

When gas accretes onto a neutron star (NS) in a low-mass X-ray binary
(LMXB), it is compressed and heated as it accumulates on the surface,
leading to thermonuclear reactions.  In many NS LMXBs, the reactions
occur unsteadily and cause Type I X-ray bursts (Grindlay et al. 1976).
Type I bursts have been observed in a large number of NS LMXBs (see
Lewin, van Paradijs \& Taam 1993; Strohmayer, Swank \& Zhang 1998 for
reviews), and the theory of these bursts is relatively well understood
(Hansen \& van Horn 1975; Woosley \& Taam 1976; Joss 1977; Taam \&
Picklum 1978;
Paczy\'nski 1982, hereafter P82; Fujimoto et al. 1984, 1987; Fushiki \&
Lamb 1987, hereafter FL87; Taam, Woosley \& Lamb 1996; Bildsten 1998).

Among the over 100 LMXBs known in the Galaxy, roughly a dozen systems,
most of them transient sources (Tanaka \& Shibazaki 1996), have been
identified as black hole (BH) candidates.  In these BH LMXBs,
dynamical measurements give mass estimates for the accreting stars
greater than the likely maximum mass $\sim3M_\odot$ of a NS (Shapiro
\& Teukolsky 1983, hereafter ST83; Narayan, Garcia \& McClintock 2001;
and references therein).

No Type I burst has been seen in a BH LMXB, even though, as we show in
this {\it Letter}, BH LMXBs ought to produce bursts as efficiently as
NS LMXBs if the accreting BH candidates possess surfaces.  The lack of
bursts thus represents possible evidence for the presence of event
horizons.  In \S2 of this paper, we examine the stability of nuclear
burning on the surface of a compact star.  In \S3, we discuss the
predictions of the model for accretion onto a $1.4M_\odot$ NS and a
$10 M_\odot$ BH candidate with a hypothetical surface.  We conclude in
\S4 with a discussion.

\section{The Model}

We consider a compact spherical star of mass $M$ and radius $R$,
accreting gas steadily at a rate $\dot\Sigma ~({\rm
g\,cm^{-2}\,s^{-1}})$.  In the local frame, the gravitational
acceleration is $g=GM(1+z)/R^2$, where the redshift $z$ is given by
$1+z=(1-R_S/R)^{-1/2}$, and $R_S=2GM/c^2$ is the Schwarzschild radius.
We assume that the accreting material has mass fractions $X_0$, $Y_0$
and $Z_0=1-X_0-Y_0$, of hydrogen, helium and heavier elements (mostly
CNO).

We consider a layer of accreted material of surface density
$\Sigma_{max}$ sitting on top of a substrate of fully burnt material
($X=Y=0$, $Z=1$).  (The particular composition of the substrate is
unimportant since we apply the inner boundary condition at its top.)
Since the physical thickness of the accreted layer is much less than
the radius, we work in plane parallel geometry and take $g$ to be
independent of depth.  We solve for the density $\rho$, the
temperature $T$, the outgoing flux $F$, and the hydrogen, helium and
heavy element fractions, $X$, $Y$, $Z=1-X-Y$, as functions of the
column density $\Sigma$ ($0\le\Sigma\le\Sigma_{max}$).

The evolution equations for the gas in the layer are the standard
equations of stellar physics.  H- and He-burning give 
\beq 
{dX\over dt}=-{\epsilon_H\over E_H^*}, \quad {dY\over
dt}={\epsilon_H\over E_H^*} -{\epsilon_{He}\over E_{He}^*}, 
\eeq 
where for our problem the total time derivative takes the form $d/dt
\equiv \partial/\partial t + \dot\Sigma\partial/\partial\Sigma$.
Here, $\epsilon_{H,He}$ are the respective nuclear energy generation
rates, and $E_{H,He}^*$ are the corresponding energy release per unit
mass of H and He burned (P82).  For $\epsilon_H$, we include the pp
chain and the CNO cycle, including fast-CNO burning, saturated CNO
burning, and electron capture reactions, as described in Mathews \&
Dietrich (1984) and Bildsten \& Cumming (1998).  Since we are not
concerned with modelling the bursts themselves, and since our
stability criterion does not depend on the detailed treatment of the
deep crust, we do not include proton captures onto heavier nuclei;
Schatz et al. (1999) illustrate some of the consequences of the
$rp$-process burning on accreting neutron stars.  For He-burning, we
include the triple-$\alpha$ reaction, but not pycnonuclear reactions
(e.g., ST83).  We do not correct the reaction rates to include
screening (e.g. FL87) since we are concerned only with determining
whether nuclear burning of H and He can proceed stably under given
conditions; stable burning of H and He utilizes almost exclusively the
reactions included.

Hydrostatic equilibrium gives $\partial P/\partial\Sigma=g$.  For the
pressure $P$ we use the expressions given in P82 for the gas,
radiation and degeneracy pressure, along with a correction when the
degenerate electrons become relativistic.  Radiative transfer gives
$\partial T/\partial\Sigma=(3\kappa F)/(16 \sigma T^3)$, where
$\sigma$ is the Stefan-Boltzmann constant and we write the opacity
$\kappa$ in the form $1/\kappa = 1/\kappa_{rad} + 1/\kappa_{cond}$. We
employ Iben's (1975) fitting functions for the radiative opacity
$\kappa_{rad}$, and an analytical formula from Clayton (1968),
suitably modified for relativistic electrons, for the conductive
opacity $\kappa_{cond}$; the latter expression agrees well with more
modern treatments (e.g. Heyl \& Hernquist 2001).  Finally, the energy
equation gives 
\beq 
\rho T{ds\over dt} = \rho(\epsilon_H+\epsilon_{He}) + \rho{\partial
F\over \partial\Sigma}, 
\eeq 
where $s$ is the entropy per unit mass.  The above five equations form
a closed set.

We have four outer boundary conditions at the surface of the star
($\Sigma=0$).  Two of these are $(X,Y)=(X_0,Y_0)$.  We obtain the
third boundary condition by equating the accretion luminosity of the
infalling gas, $L_{acc}=4\pi R^2\dot\Sigma c^2z/(1+z)$, to blackbody
emission from the surface: $4\pi R^2\sigma T_{out}^4=L_{acc}$.  This
gives the surface temperature $T_{out}$.  Then, using $T_{out}$ and an
assumed value of $F_{out}$, we solve for the surface density profile
$\rho(\Sigma)$ from the radiative transfer equation, thus obtaining
the fourth boundary condition.

At the base of the accreted layer we have an inner boundary condition.
We assume that the temperature at the top of the substrate,
$T_{in}=T(\Sigma_{max})$, is fixed.  We examine several values of
$T_{in}$ for layers with $\Sigma_{max}=10^9, 10^{10}$ and $10^{11}$~g
cm$^{-2}$.  Applying the boundary condition at $\Sigma_{max}$ rather
than deeper down is an approximation, but the error due to this is not
serious.  For the high surface densities we consider, the heat
transfer is dominated by conduction (Heyl \& Hernquist 2001), so the
temperature gradient for $\Sigma>\Sigma_{max}$ is small.  Moreover, we
examine models for several values of $T_{in}$ which further mitigates
any error.

The calculations proceed in two stages.  First, we solve for the
steady state profile of the accretion layer by making the replacement
$d/dt \to \dot\Sigma d/d\Sigma$, $\partial/\partial\Sigma \to
d/d\Sigma$ in the governing equations.  This gives five ordinary
differential equations with four outer boundary conditions and one
inner boundary condition.  The solution gives the profiles of the
basic fluid quantities: $X(\Sigma), ~Y(\Sigma), ~P(\Sigma)=g\Sigma,
~\rho(\Sigma), ~T(\Sigma)$.

Having calculated the steady state structure of the accretion layer,
we next check its stability.  Various local stability criteria have
been discussed in the literature (e.g., Bildsten 1998), in which one
considers the properties of the gas at a single depth.  FL87 proposed
a global criterion involving an integral over the entire layer.  While
an improvement, this approach is still unsatisfactory since the
authors assumed a constant temperature perturbation throughout the
layer.

We have carried out a full linear stability analysis of the accretion
layer.  We start with the steady state solution and assume that it is
slightly perturbed, $Q(\Sigma) \to Q(\Sigma) +Q'(\Sigma)\exp(\gamma
t)$, where $Q$ corresponds to each of our five variables, and the
perturbations $Q'(\Sigma)$ are taken to be small.  We linearize the
five equations described earlier (three of which include time
derivatives), apply the boundary conditions, and solve for the
eigenvalue $\gamma$.  We obtain a large number of solutions for
$\gamma$ (technically, there is an infinite number since we are
dealing with continuous functions), many of which have both a real and
imaginary part.  We consider the accretion layer to be unstable if any
eigenvalue $\gamma$ has a real part (growth rate) greater than the
characteristic accretion rate $\gamma_{acc}=\dot\Sigma/\Sigma_{max}$.

If the steady-state model is unstable according to the above
criterion, it cannot accrete matter stably with the particular
$\Sigma_{max}$ and $\dot\Sigma$.  Whether this instability manifests
itself as a Type~I burst depends on how the burning flame, once
ignited, envelopes the surface of the star.  This is as yet an
unsolved problem, although Spitkovsky, Levin \& Ushomirsky (2002)
present a possible solution.  In the following we assume that the
instability will grow rapidly to the nonlinear regime and that the
system will indeed exhibit Type I bursts.

\section{Results}

Figure 1 shows results for solar composition material ($X_0=0.7$,
$Y_0=0.27$, $Z_0=0.03$) accreting on two kinds of compact stars.

\clearpage
\begin{figure*}
\plotone{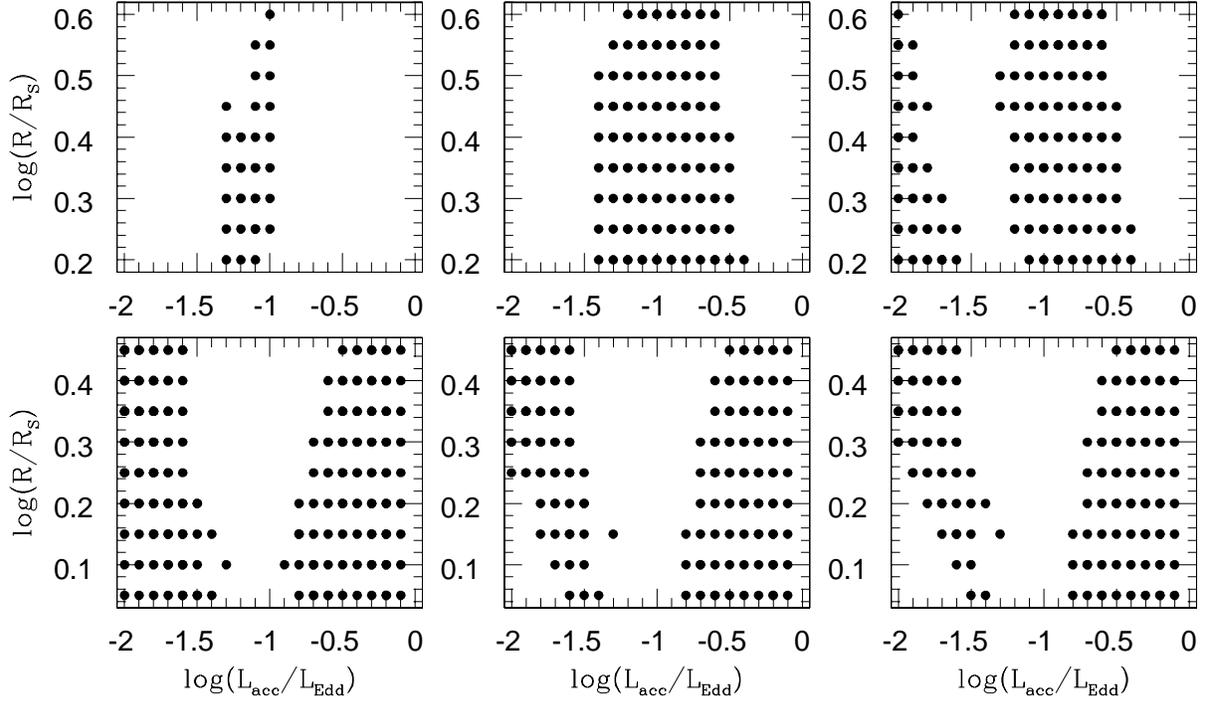}
\caption[f1.eps]{Regions of instability, shown by
dots, as a function of accretion luminosity and stellar radius.  Top
left: $1.4M_\odot$ NS with a base temperature $T_{in}=10^{8.5}$~K.
Top center: $T_{in}=10^8$K.  Top right: $T_{in}=10^{7.5}$ K.  Bottom
left: $10M_\odot$ BH candidate with a surface, and a base temperature
$T_{in}=10^{7.5}$~K.  Bottom center: $T_{in}=10^7$ K.  Bottom right:
$T_{in}=10^{6.5}$ K.}
\label{fig2}
\end{figure*}
\clearpage

The top three panels correspond to an accreting NS of mass
$1.4M_\odot$.  We consider a range of accretion rates, parameterized
by the ratio $L_{acc}/L_{Edd}$, where we take the Eddington luminosity
to be $L_{Edd}=4\pi GMc/\kappa_{es}$ with $\kappa_{es}=0.4 ~{\rm
cm^2\,g^{-1}}$.  We also consider a range of radii for the NS:
$\log(R/R_S)=0.2-0.6$ ($R=6.5-16$ km).  For each choice of
$L_{acc}/L_{Edd}$ and $R/R_S$, we try three values of the surface mass
density of the accreted layer: $\Sigma_{max} = 10^9, ~10^{10},
~10^{11} ~{\rm g\,cm^{-2}}$.  If any of the three cases is unstable,
i.e., if it has any eigenvalue with ${\rm Re}(\gamma) >\gamma_{acc}$,
then we claim that the system will exhibit instabilities that may grow
into Type~I bursts.

The results presented correspond to three choices of the temperature
at the base of the accreted layer: $T_{in}=10^{8.5}, ~10^8$ and
$10^{7.5}$~K from left to right.  In this {\it Letter} we are
primarily interested in transient LMXBs.  Since these sources have
very low luminosities in quiescence ($L_X<10^{33} ~{\rm
erg\,s^{-1}}$), the core temperatures of the NSs are likely to be
$T_{in}\la10^8$ K (Possenti et al.  2001; 
Brown, Bildsten \& Chang 2002).

The calculations shown in Fig. 1 indicate that NSs are unstable to
bursts for a wide range of $T_{in}$, but that the width of the
instability strip (as a function of $L_{acc}$) is less for higher
values of $T_{in}$.  The reason for this is clear from the analysis of
P82 who showed that when the flux escaping from the stellar core into
the accretion layer increases (which happens when $T_{in}$ increases),
bursting behavior is restricted to a smaller range of $\dot \Sigma$.

In the case of transient NS LMXBs, we expect $T_{in}$ to be $\la10^8$
K.  Fig. 1 indicates that these sources should be unstable to bursts
for accretion luminosities up to $L_{acc} \approx 0.3 L_{Edd}$.  The
predicted luminosity limit is generally consistent with observations;
the only NSs that are known not to burst are bright Z sources with
$L_{acc}\to L_{Edd}$ (e.g., Matsuba et al. 1995), and X-ray pulsars.
Although the latter are significantly less luminous than Eddington,
they accrete effectively at close to the Eddington rate since the
accreting matter is channeled onto a small area on the NS surface by
strong magnetic fields (see Lamb 2000 for a detailed discussion of
this argument).

Below $L_{acc}\sim0.3L_{Edd}$, the instability is initially of a mixed
form in which a He-burning instability triggers a burst in which H and
He both burn explosively.  At lower luminosities, nearly all the H is
burned steadily and the instability corresponds to a pure He burst.
These results are consistent with previous work (e.g., Bildsten 1998).
The calculations indicate that bursting behavior cuts off below an
accretion luminosity $L_{acc}\sim10^{-1.5}L_{Edd}$.  The cutoff is the
result of the restriction $\Sigma_{max}\le10^{11} ~{\rm g\,cm^{-2}}$
in our models.  Systems with luminosities below our cutoff are still
unstable to He bursts, but only at extremely high column density
$\Sigma_{max}$.  We have ignored such bursts since the recurrence time
$t_{rec}$ is too long to be of interest for observations of transient
X-ray binaries.

For $T_{in}=10^{7.5}$ K, there is a second instability strip at low
luminosities $L_{acc}\sim10^{-2}L_{Edd}$.  This strip corresponds to
pure H bursts; the possibility of such bursts was first noted by
Fujimoto et al. (1987).  An interesting difference between the two
instability strips is that the strip on the right generally has
complex eigenvalues $\gamma$ for the unstable modes while that on the
left has real eigenvalues.  As explained above, the gap between the
two strips is real if we restrict ourselves to reasonable values of
$\Sigma_{max}$ and $t_{rec}$.  We are not aware of clear observational
evidence for or against the gap, but such evidence could be searched
for in future observations.

The bottom three panels in Fig. 1 show results for a $10M_\odot$ BH
candidate with a surface.  The three panels correspond to different
choices of the base temperature: $T_{in}=10^{7.5}, ~10^7$ and
$10^{6.5}$~K from left to right.  The particular choices of $T_{in}$
are motivated by the extraordinarily low quiescent luminosities of
many transient BH LMXBs ($L_X<10^{31} ~{\rm erg\,s^{-1}}$, Narayan et
al. 2001; a few BH LMXBs are brighter than this limit, but even these
are not likely to have $T_{in}\ga10^{7.5}$ K).  We consider stellar
radii in the range $\log(R/R_S)=0.05-0.45$, corresponding to $R$
between $(9/8)R_S$ and about $3R_S$, a factor of nearly twenty in the
surface gravity.  The choice $(9/8)R_S$ corresponds to the smallest
radius in general relativity for an object whose density either
decreases or remains level with increasing radius (ST83).

The calculations indicate that BH candidates with surfaces are not
very different from NSs in their bursting behavior.  Except for a
modest rightward shift of the positions of the instability zones, the
results in the lower three panels are quite similar to those for a NS
with a similarly low value of $T_{in}$ (upper right panel).  As in the
case of NSs, bursts are expected for BH candidates even for larger
values of $T_{in}$ than we have considered, except that the
instability strip becomes narrower when $T_{in}$ exceeds $10^8$ K.  We
conclude that BH candidates are as prone to the instabilities that
lead to Type I bursts as NSs are.  The absence of bursts in BH LMXBs
is thus highly significant and argues for the lack of surfaces in
these systems.

Although our analysis is for a spherical, non-rotating $10M_\odot$
star, the results are not expected to differ greatly for a rotating
object.  The effective surface gravity and accretion rate vary by only
a factor of 2--3 from pole to equator even for a maximally rotating
oblate ellipsoid that is on the verge of the secular triaxial
instability; furthermore, the dependence on compactness and equation
of state is small (see Gondek-Rosinska \& Gourgoulhon 2002; ST83).  In
comparison, Fig. 1 shows that bursts are present for a wide range of
surface gravity $g$ (factor of 20) and accretion rate $\dot\Sigma$
(factor of 100).  The rather modest variation of $g$ and $\dot\Sigma$
with latitude in a rotating star is thus not likely to have an effect.

Because our model focuses only on the most important physical effects,
and neglects many details, the exact positions of the instability
strips in Fig. 1 may be uncertain at the level of say a factor of two
in accretion luminosity.  We believe, however, that the overall
pattern of instability we have computed is fairly robust.  Also, the
calculations we present here do not directly predict what kind of
bursts are produced.  We leave this for a more detailed paper, but
briefly, we find that for NSs with $L_{acc}\sim0.1L_{Edd}$, bursts
have durations of a few seconds and $t_{rec}\sim10$ hours.  For lower
$L_{acc}$, as is well-understood, bursts have longer $t_{rec}$ and larger
fluences.  These results agree qualitatively with observations.

\section{Discussion}

It is clear from the theory of bursts (Bildsten 1998, and references
therein) that bursting behavior is largely determined by the surface
gravity $g$, the mass accretion rate, and the composition of the
accreting material.  Since these parameters are similar in NS and BH
systems, the bursting behavior of the two should be similar.  We have
quantified this argument with a global linear stability analysis which
represents a technical advance over previous calculations.  

The results presented in Fig. 1 show that if BH candidates had
surfaces they ought to experience thermonuclear instabilities as
commonly as NSs do; by inference, they should have frequent Type~I
bursts.  However, no BH candidate has exhibited Type I bursts.  The
most obvious explanation is that NSs have surfaces and BH candidates
have event horizons (Menou 2001).  If there is no surface, material
cannot accumulate, and therefore cannot become hot or dense enough to
trigger a thermonuclear explosion.

Before we can claim that this ``proves'' the reality of the event
horizon, more work is needed.  First, we need to show that the model
is able to reproduce the more detailed features of Type I bursts as
observed in NS LMXBs.  The statistics of burst durations and
recurrence times (e.g., van Paradijs, Penninx \& Lewin 1988) and the
occurrence of oscillatory behavior in some systems (Revnivtsev et
al. 2000) ought to appear naturally in a realistic model.  Also, any
NS systems that burst when they should not by our calculations, or do
not burst when they should according to the model, need to be
explained.  Second, the role of the inner boundary condition needs to
be explored in detail.  In Fig. 1, we see that different choices of
the base temperature for a BH candidate give similar results.  We have
tried other boundary conditions, and also tried changing the
composition of the accreting gas.  In all cases we find that the
accumulating layer is unstable to bursts for a wide range of
luminosity, both in $1.4M_\odot$ NSs and $10M_\odot$ compact stars
with surfaces.  Finally, the difficult issue of flame propagation over
the surface of the star once the instability has been triggered needs
to be addressed (e.g., Lamb 2000, Spitkovsky et al. 2002); the effect
of rotation, for instance, is presently unclear.

We should caution that the analysis presented here assumes that the
accumulating gas on the surface of a BH candidate behaves like normal
matter, with nucleons and electrons; this is the case for standard
models of strange stars and Q-stars (Glendenning 1997).  Obviously,
the argument becomes invalid if the properties of the gas change
drastically, e.g., if the nuclei disappeared and were replaced by
quarks.  Whether such extreme changes are plausible remains to be
seen.  The density and pressure at the base of the bursting layer do
not go above ${\rm few}\times10^8 {\rm g\,cm^{-3}}$ and $10^{26} ~{\rm
erg\,cm^{-3}}$ even in the most extreme cases we have considered.  It
is hard to imagine exotic physics being important under these
conditions (Glendenning 1997).

On the observational front, we should check whether some NSs that lie
within the unstable regions of Fig. 1 are stable to bursts.  Any
obvious large-scale disagreement between the observed burst behavior
of NS LMXBs and the results presented here would indicate that the
model is missing important physics.  In the case of BH binaries, we
should use observational data to derive quantitative limits on
bursting activity.  The transient BH LMXBs are particularly important
for such work since they vary over a wide range of $L_{acc}$ during
their accretion outbursts (Tanaka \& Shibazaki 1996).  If these
sources have surfaces it is virtually impossible to arrange matters
such that the objects have no bursts at all over the entire range of
$L_{acc}$.  A firm demonstration that BH transients do not have Type I
bursts would thus be a strong argument for the presence of event
horizons in these systems.

\acknowledgements 

We thank Andrew Cumming, Alex Ene, Kristen Menou, Bohdan Paczy\'nski and
Greg Ushomirsky for useful discussions, and two referees for useful
comments.  RN was supported in part by the W. M. Keck Foundation as a
Keck Visiting Professor.  RN's research was supported by NSF grant
AST-9820686 and NASA grant NAG5-10780.  JSH was supported by the
Chandra Postdoctoral Fellowship Award \# PF0-10015 issued by the
Chandra X-ray Observatory Center, which is operated by the Smithsonian
Astrophysical Observatory for and on behalf of NASA under contract
NAS8-39073.

\end{document}